\documentclass[showpacs,twocolumn,prb,superscriptaddress]{revtex4-1}
\usepackage{amsfonts}
\usepackage{graphicx}
\usepackage{amsmath}
\usepackage{amssymb}
\usepackage{color}

\def\bea {\begin{eqnarray}}
\def\eea {\end{eqnarray}}
\def\be {\begin{equation}}
\def\ee {\end{equation}}
\def\up{\uparrow}

\DeclareMathOperator{\imag}{Im}
\DeclareMathOperator{\real}{Re}

\begin{document}

\title{The BCS Pairing Instability in the Thermodynamic Limit}
\author{F.\ Marsiglio}
\author{K.\ S.\ D.\ Beach}
\affiliation{Department of Physics, University of Alberta, Edmonton, Alberta, Canada, T6G~2E1}
\author{R.\ J.\ Gooding}
\affiliation{Department of Physics, Engineering Physics and Astronomy, Queen's University, Kingston ON K7L~3N6}

\begin{abstract}
The superconducting pairing instability---as determined by a divergence of the two-particle susceptibility---is obtained in the mean field (BCS) approximation in the thermodynamic limit. The usual practice is to examine this property
for a finite lattice. We illustrate that, while the conclusions remain unchanged, the technical features are
very different in the thermodynamic limit and conform more closely with the usual treatment of phase transitions encountered
in, for example, the mean-field paramagnetic-ferromagnetic transition. Furthermore, by going to the extreme dilute limit, one
can distinguish three dimensions from one and two dimensions, in which a pairing instability occurs even for two particles. 
\end{abstract}


\date{\today}
\maketitle

\section{Introduction}

The Bardeen, Cooper, and Schrieffer (BCS) theory of superconductivity, following the original literature,\cite{bardeen57} 
is typically first presented in textbooks as a proposed variational ground state, whose energy is lower than that of the 
corresponding normal state.\cite{schrieffer64,rickayzen65} This is followed by a discussion of the excited states, from 
which, in weak coupling at least, a critical temperature is derived, corresponding to the breakup of pairs.

An alternate view of the transition was put forward by Thouless,\cite{thouless60} who tracked the BCS instability
from above (in temperature) by monitoring the two particle propagator. By including a specific set of processes (denoted by ``ladders'' in 
the particle-particle channel of the
diagrammatic version of this formulation), one finds an instability of the normal phase. This approach
 is also explained in many texts\cite{schrieffer64,ambegaokar69} and will be briefly summarized below.

Seeing the superconducting transition as an instability of the normal state at some finite temperature highlights the
potential importance of pairing fluctuations that occur in the normal state even before the critical transition is reached. The possibility
of pairing fluctuations has been important in the elucidation of the so-called pseudo-gap that occurs in the high-$T_c$ cuprate
materials. One school of thought regards the pseudo-gap in these materials as a tell-tale signature of pairing fluctuations in the normal 
state.\cite{timusk99} 

There is now an extensive literature on the presence of pairing fluctuations in the normal state and on their impact on various normal 
state properties.\cite{pairing_review,gooding04,chen05} However, our purpose here is to revisit the simple so-called Thouless criterion 
for the BCS instability and to reformulate it in the thermodynamic
limit. We have always found it peculiar that this instability is signalled by the appearance of two {\em imaginary} roots in the denominator of the two-particle propagator, whereas other instabilities appear to be accompanied by the appearance of a {\em real} root.\cite{doniach74} It turns out that the two-particle pairing instability is always, to our knowledge, formulated for a finite system; in the thermodynamic limit, as we show below, the 
criterion behaves quite differently and much more like other instabilities in condensed matter.

For simplicity we focus on the simplest model that exhibits superconductivity (at least at the mean field level), the attractive Hubbard model
on a hypercubic lattice in one, two, and three dimensions.\cite{beach01}
While considerable attention has been devoted to single-particle properties, since these are often
most related to the measured properties,\cite{timusk99} the two-particle properties are the ones
that are key to understanding single-particle properties.\cite{vilk97} For our purposes, we express the two-particle propagator
in the non-self-consistent ladder approximation,\cite{beach01}
\def\xq{\tilde{\chi}_0({\bf q},i\nu_n)}
\def\fact{1 \over N \beta}
\be
g_2({\bf q},i\nu_n) \equiv {\chi_0({\bf q},i\nu_n) 
\over 1 - |U| \chi_0({\bf q},i\nu_n)},
\label{pairq}
\ee
where $\chi_0$ is the ``noninteracting'' pair susceptibility
\be
\chi_0({\bf q},i\nu_n) = {1 \over N \beta} \sum_{{\bf k},m} G_{0 \up}({\bf k},i\omega_m) 
G_{0 \downarrow}({\bf q} - {\bf k},i\nu_n-i\omega_m ),
\label{susc}
\ee 
and $G_{0 \sigma}({\bf k},i\omega_m)  = 
[i\omega_m - (\epsilon_{\bf k} - \mu)]^{-1}$
is the noninteracting single-particle propagator.
Here, {\bf k} and {\bf q} are wave vectors, and $\epsilon_{\bf k}$ is the single electron dispersion appropriate to tight-binding
with nearest neighbour hopping. The Matsubara frequencies are defined as
$i\omega_m \equiv \pi T (2m - 1)$ for Fermions and $i\nu_n \equiv i2\pi T n$ for Bosons.
$\beta \equiv (k_B T)^{-1}$ is the inverse temperature, and $N$ is the number of lattice sites.
All wavevector summations span the entire Brillouin zone, and Matsubara sums go over all
integers.

For the BCS instability we can focus on ${\bf q} = 0$ and, in fact, $\nu_n = 0$; nonetheless, we wish to illustrate the instability by monitoring 
$g_2$ as a function of (real) frequency. 
The result for $\chi_0$ is
\be
\chi_{0}({\bf q},z) = - {1 \over N} \sum_{\bf k} {1 - f(\epsilon_{\bf k} - \mu) -
f(\epsilon_{-{\bf k} +{\bf q}} - \mu) \over z - (\epsilon_{\bf k} - \mu) -
(\epsilon_{-{\bf k} +{\bf q}} - \mu)},
\label{susc00}
\ee
where we have now analytically continued the result to the upper half-plane 
($i\nu_n \rightarrow z$), and in particular for $z = \nu + i\delta$, with $\delta$ a positive infinitesimal. Here $f(x) \equiv 1/(e^{\beta x} + 1)$ is the Fermi-Dirac distribution function. Equation~\eqref{susc00} is the one displayed in textbooks and reviews. 
We show its real and imaginary parts in Fig.~1, for a finite (1D) system with $N = 32$ at half-filling ($\mu = 0$)
and for a number of temperatures. The chemical potential can either be held fixed or tuned to produce a desired electron density, $n$.
\begin{figure}[tp]
\begin{center}
\includegraphics[height=9cm,width=5cm,angle=-90,scale=0.9]{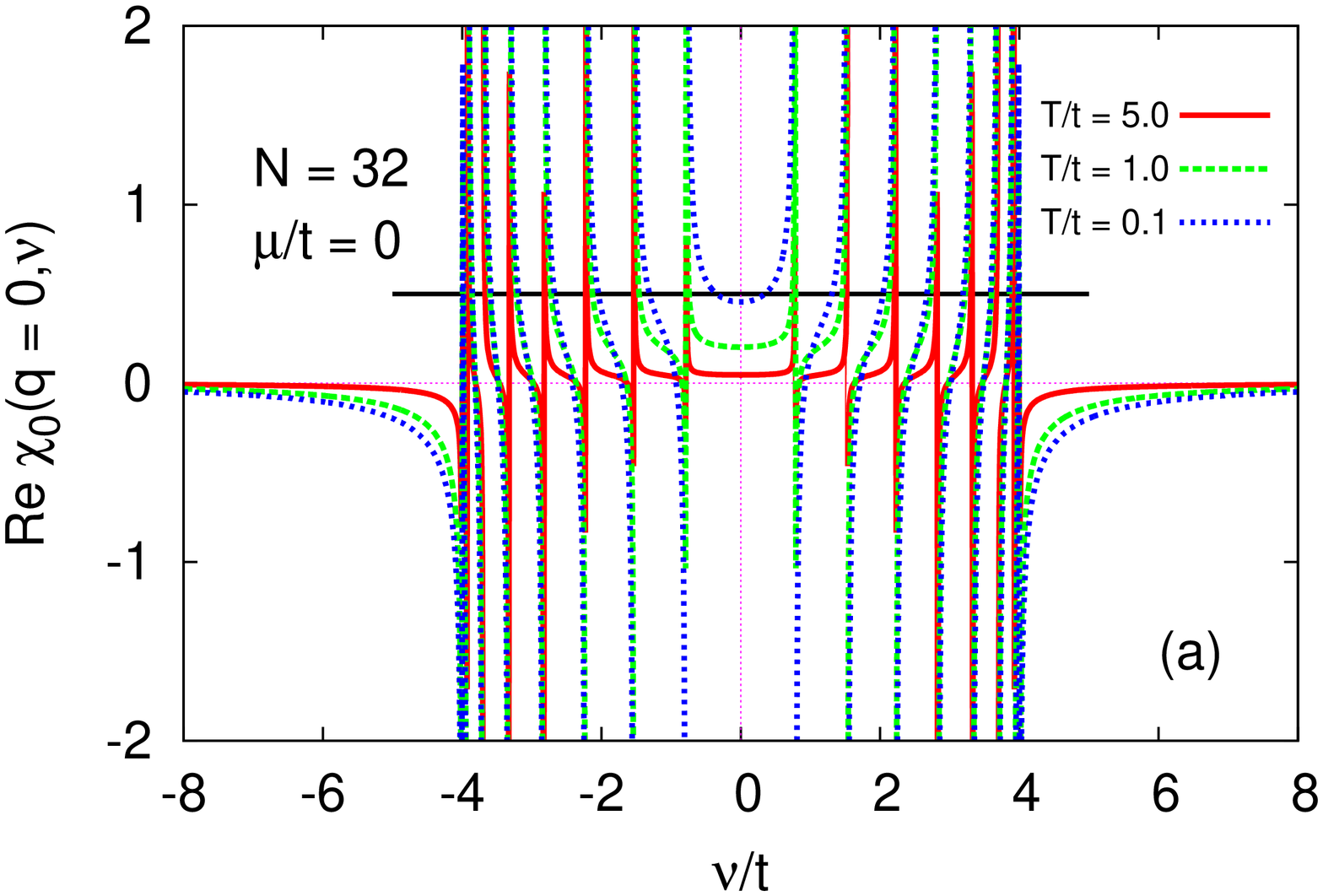}
\includegraphics[height=9cm,width=5cm,angle=-90,scale=0.9]{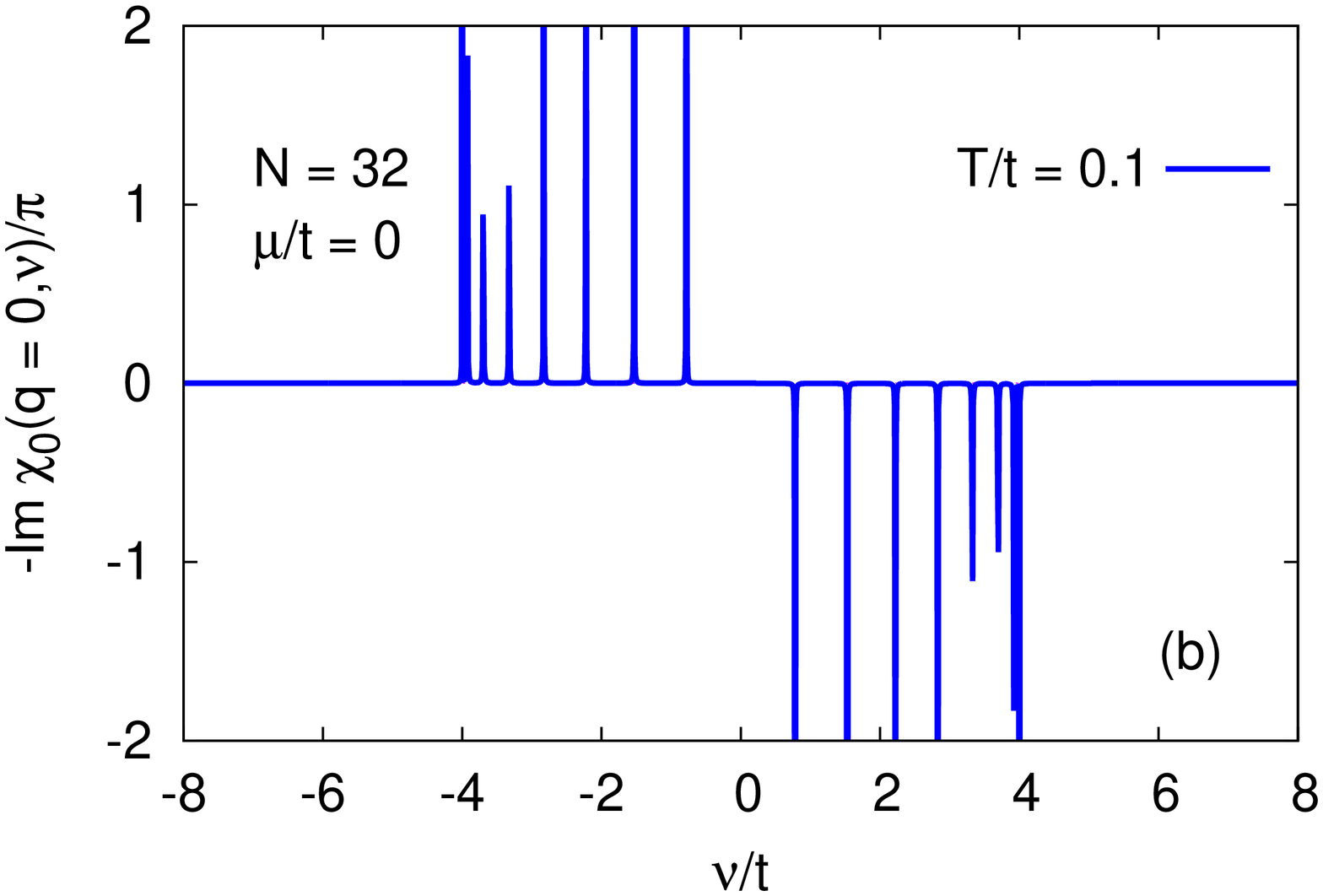}
\end{center}
\caption{ Real (a) and Imaginary (b) parts of the noninteracting pair susceptibility
at zero wave vector vs.\ frequency, for three different temperatures. Note that
poles occur at the energies corresponding to two single electron energies. In
addition, the minimum at zero frequency diverges as the temperature goes to
zero (not evident in (a) because the divergence is logarithmic). The figures were
produced in one dimension with a finite lattice of length 32 sites. The horizontal
line at $0.5$ in (a) denotes the value of $1/|U|$ for $|U|=2t$. As the central minimum rises
above this horizontal line (with decreasing temperature) an instability is signalled, as discussed
in the text.}
\label{fig1}
\end{figure}

The use of a finite lattice necessarily produces a series of poles, as is evident in Fig.~1(b), where the imaginary part of $\chi_0$ is plotted (a small numerical value of
$\delta$ is used in place of an infinitesimal). These poles correspond to the noninteracting case ($|U|=0$); 
they arise from roots of the denominator in Eq.~\eqref{susc00} with a small {\em negative} imaginary part and hence
lie in the lower half plane. This fact renders them ``innocent,'' since, when substituted into the original time-dependent propagator, $e^{-i\nu t}$, they will decay away with time. They correspond to normal two-particle excitations and result in a density 
of states that extends across the two-particle continuum. In this one dimensional case, these energies extend from 
twice the energy of the bottom of the band ($-4t$) to twice the energy of the top of the band ($4t$). At high temperatures
the excitation energies occur at energies very close to those of the noninteracting case, indicated by the vertical asymptotes in Fig.~1(a); in this case the energies are all slightly shifted by the interaction and occur not at the asymptotes but at the intersection of the curve with the horizontal black line, representing $1/|U|$.
These intersections correspond to the zeros of the denominator in Eq.~\eqref{pairq} and will be referred to as the poles of the 
two-particle propagator.

A special case, with a distinctive temperature dependence, is the pair of zeros near $\nu = 0$.
As the temperature decreases, the
zero frequency minimum in Fig.~1(a) increases in value and eventually crosses $1/|U|$.
As it does so, two real roots (with small negative imaginary part) become two pure imaginary roots, one of which is 
in the upper half plane. Substitution of this root into $e^{-i\nu t}$ results in an excitation that blows up with time, indicative of an unstable 
normal phase.\cite{kadanoff61,schrieffer64,ambegaokar69}

We now outline the situation in the thermodynamic limit, which, while giving the same physics, looks quite different.
For ${\bf q} = 0$ the discrete sum in Eq.~\eqref{susc00} is converted to an integral over the single-particle density of states, $g(\epsilon)$:\cite{remark}
\be
\chi_0(\nu + i\delta) = -\int_{-W/2}^{+W/2} \! d\epsilon\, g(\epsilon) {\tanh\bigl( \beta (\epsilon - \mu)/2 \bigr) \over \nu + i\delta - 2(\epsilon - \mu)},
\label{susc1}
\ee
where $\pm W/2$ is the top (bottom) of the single electron band ($\pm 2t$ in 1D, $\pm 4t$ in 2D, and $\pm 6t$ in 3D), and since ${\bf q} = 0$ 
we have omitted it from the argument list for $\chi_0$. This integral requires a principal value part, which
can be done analytically:
\begin{multline} \label{susc2}
\chi_0(\nu + i\delta) = {1 \over 2} g\bigl({\nu \over 2} + \mu\bigr) \tanh{\beta \nu \over 4} \log{\Biggl\{{{W\over 2} - \mu - {\nu + i\delta \over 2} \over {W\over 2} + \mu + {\nu + i\delta \over 2}}\Biggr\}} \\
+ i{\pi \over 2} g\bigl({\nu \over 2} + \mu\bigr) \tanh{\beta \nu \over 4} \\
+{ 1 \over 2} \int_{-W/2}^{+W/2} \!d\epsilon\, {g(\epsilon) \tanh{\beta (\epsilon - \mu) \over 2} - g\bigl({\nu \over 2} + \mu\bigr) \tanh{\beta \nu \over 4}
\over \epsilon - \mu - (\nu + i\delta)/2}.
\end{multline}
The integration on the last line is no longer singular and can be computed by quadrature.

In Figs.~2(a) and 2(b) we show the corresponding results for an infinite 
system (in 1D) at temperatures $T = 1$, $0.1$, and $0.01$ (in units of 
$t$; hereafter all energies will be quoted in units of $t$). The real 
part of $\chi_{0}(\nu)$ clearly shows a {\it maximum}
at zero frequency; elsewhere there are no positive divergences as they 
have been integrated to a smooth curve in the principal value sense. 
The negative divergences occur at the band edges and are due to the 
divergent single electron density of states at the band edges
in one dimension. 

\begin{figure}
\begin{center}
\includegraphics[height=9cm,width=5cm,angle = -90,scale=0.9]{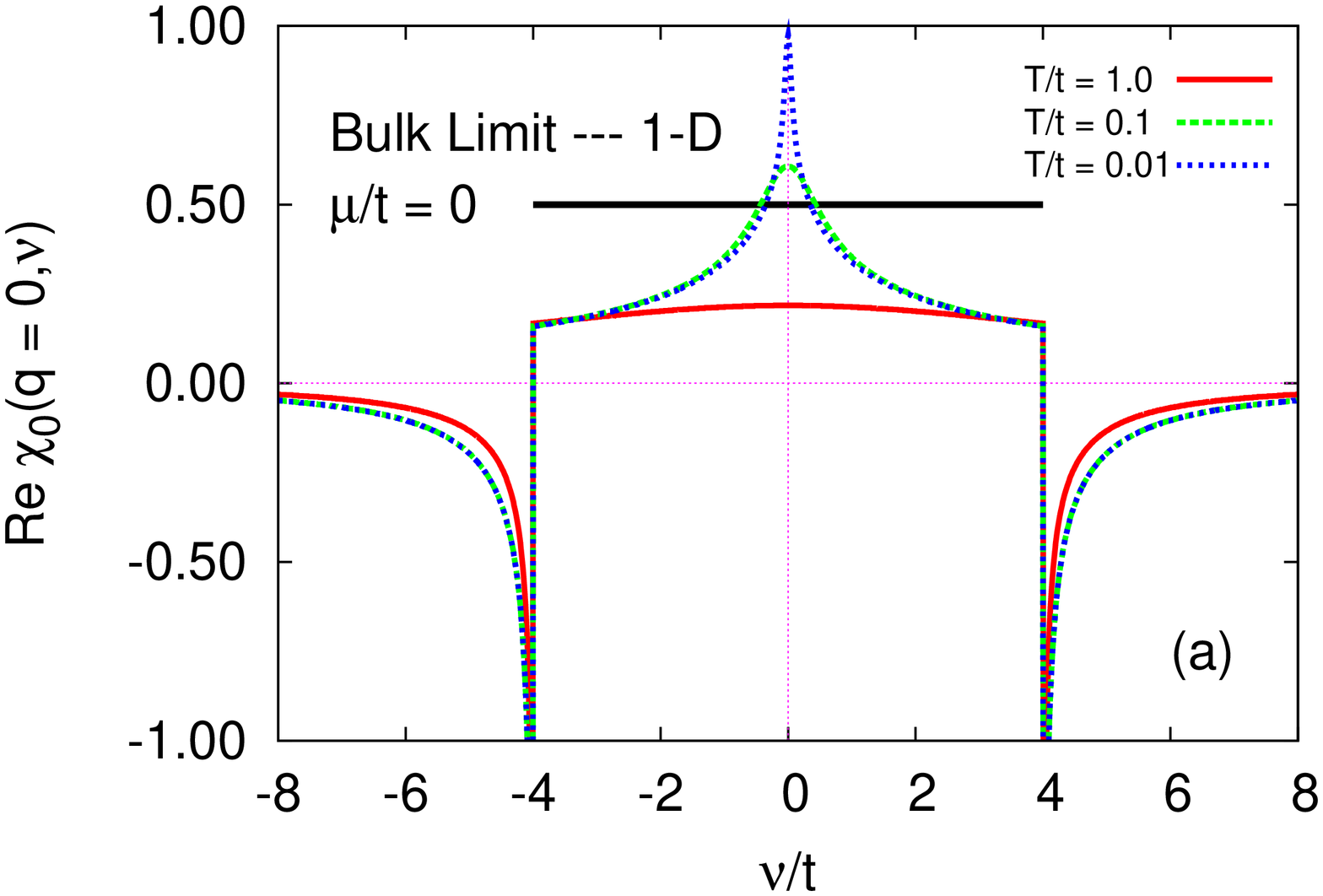}
\includegraphics[height=9cm,width=5cm,angle=-90,scale=0.9]{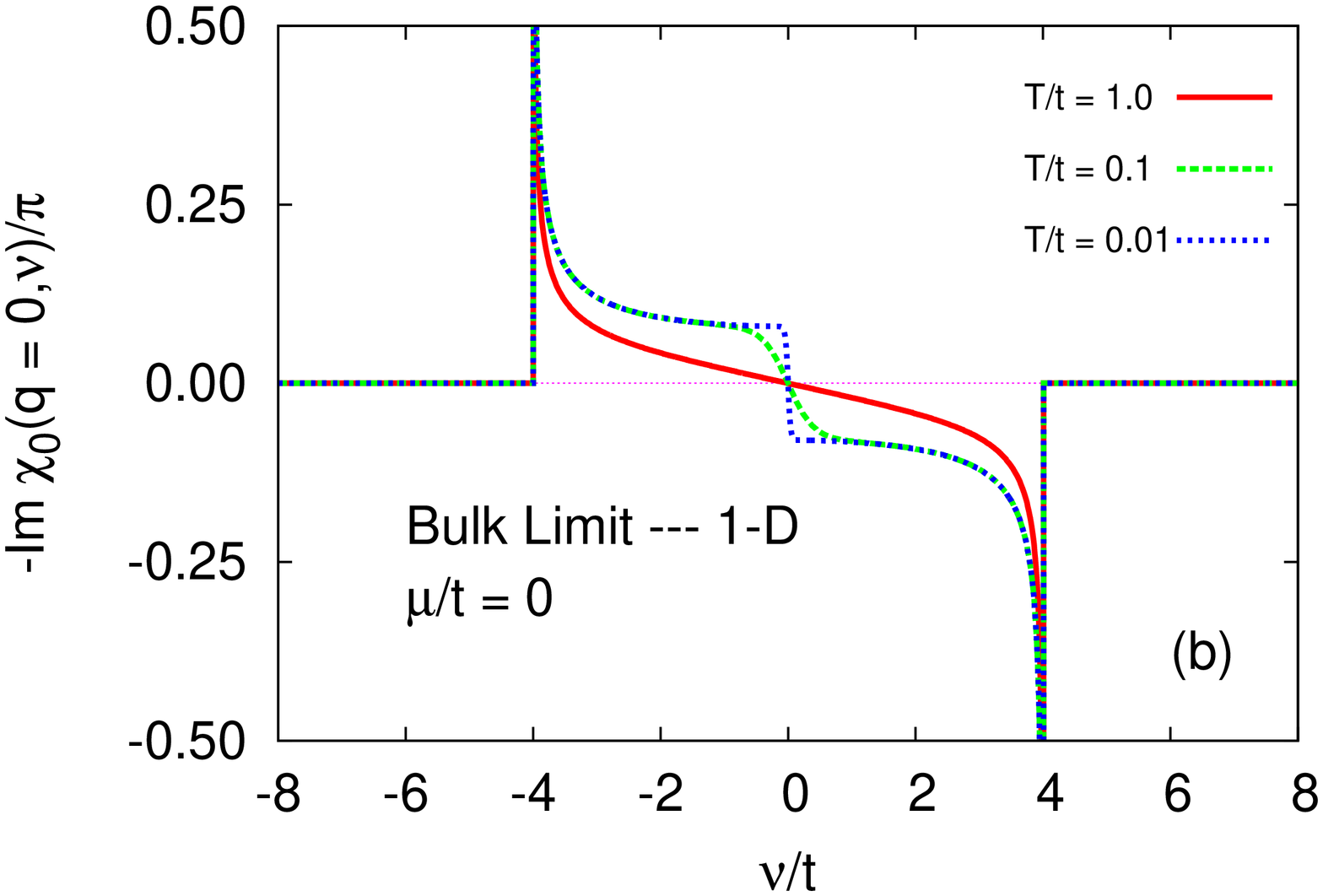}
\caption{Real (a) and Imaginary (b) parts of the noninteracting susceptibility
at zero wavevector vs.\ frequency, for three different temperatures, for the bulk
limit in one dimension. Note that part (a) in particular looks very different from
the finite size counterpart in Fig.~(1a). In particular, the poles corresponding
to sums of single electron energies are evident only in (b). The developing
singularity at zero frequency remains, as is evident in (a).}
\label{fig2}
\end{center}
\end{figure}

As is apparent from the figure, these ``band edge'' divergences are present at 
all temperatures. In fact, for the lowest two temperatures shown, the curves
are essentially the same {\it except for the region near zero frequency}, 
where the maximum diverges as $T \rightarrow 0$, indicative of
a superconducting/charge-density-wave instability. In fact at zero temperature
the real part is given analytically by the following expression (for $\mu = 0$):
\begin{equation}
\real \chi_0(\nu + i \delta) =
\begin{cases}
     {1 \over 4\pi t} {1 \over \sqrt{1 - \bar{\nu}^2} } \log{ 
     \biggl(
     {1 + \sqrt{1 - \bar{\nu}^2} \over 1 - \sqrt{1 - \bar{\nu}^2} }
     \biggr)} & 
\text{$\bar{\nu} < 1$},
\cr
     -{1 \over 2\pi t} {1 \over \sqrt{\bar{\nu}^2 - 1}} \arctan{\biggl( {1 \over
     \sqrt{\bar{\nu}^2 - 1}} \biggr)} &
\text{$\bar{\nu} > 1$},
\end{cases}
\label{analytical}
\end{equation}
where $\bar{\nu} \equiv \nu/(4t)$. The divergence at zero frequency is
evident in this expression. At finite temperature an exact analytical expression
is not possible, even for zero frequency. However, to a very good approximation
one can obtain $\real \chi_0(\nu = 0) = {1 \over 2 \pi t} \log{\bigl(
1.13 {4t \over T} \bigr)}$. Notice that the argument of the natural logarithm
is a factor of 2 larger than what would have been obtained by simply approximating
the density of states as a constant at the chemical potential ($\mu = 0$ in this case).

For a nonzero attractive interaction, an instability is signalled by the maximum crossing the black horizontal line 
positioned at $1/|U|$. This signals the onset of an instability in a way that is familiar from studies in mean-field ferromagnetism,\cite{doniach74} for example. It {\em appears}
as though two {\em real} roots are emerging; in fact a careful analysis of Eq.~\eqref{pairq}, using
$\chi_0(\nu + i\delta) \approx a_0 + ic_0 \nu$, with $a_0$ and $c_0$ positive real constants [see Figs.~2(a) and (b)]
near $\nu \approx 0$ shows that as $a_0$ traverses $1/|U|$, the same root with {\em negative} imaginary part
becomes a root with a {\em positive} imaginary part. As in the finite lattice case, then, the two particle propagator
becomes unstable in time.

Figure~(2b) shows the spectral function, $B_{0}({\bf q},\nu) \equiv -\imag 
\chi_{0}({\bf q},\nu + i \delta)/\pi$ as a function of frequency. Aside from the 
asymmetrization, this quantity provides an image of the single electron
density of states. This remains true in any dimension, as can be seen 
from taking the imaginary part of Eq.~\eqref{susc2}:
\be
B_{0}({\bf q}= 0,\nu) = -{1 \over 2} \tanh \biggl({\beta \nu \over 4} \biggr)
g({\nu \over 2} + \mu).
\label{b00}
\ee
As the temperature approaches zero, the
hyperbolic tangent function simply changes the sign of the density of states at the origin.
In Fig.~(1b) the delta function structure
was merely providing an image of the finite system's discretized density of states.
\begin{figure}
\begin{center}
\includegraphics[height=9cm,width=5cm,angle=-90,scale=0.9]{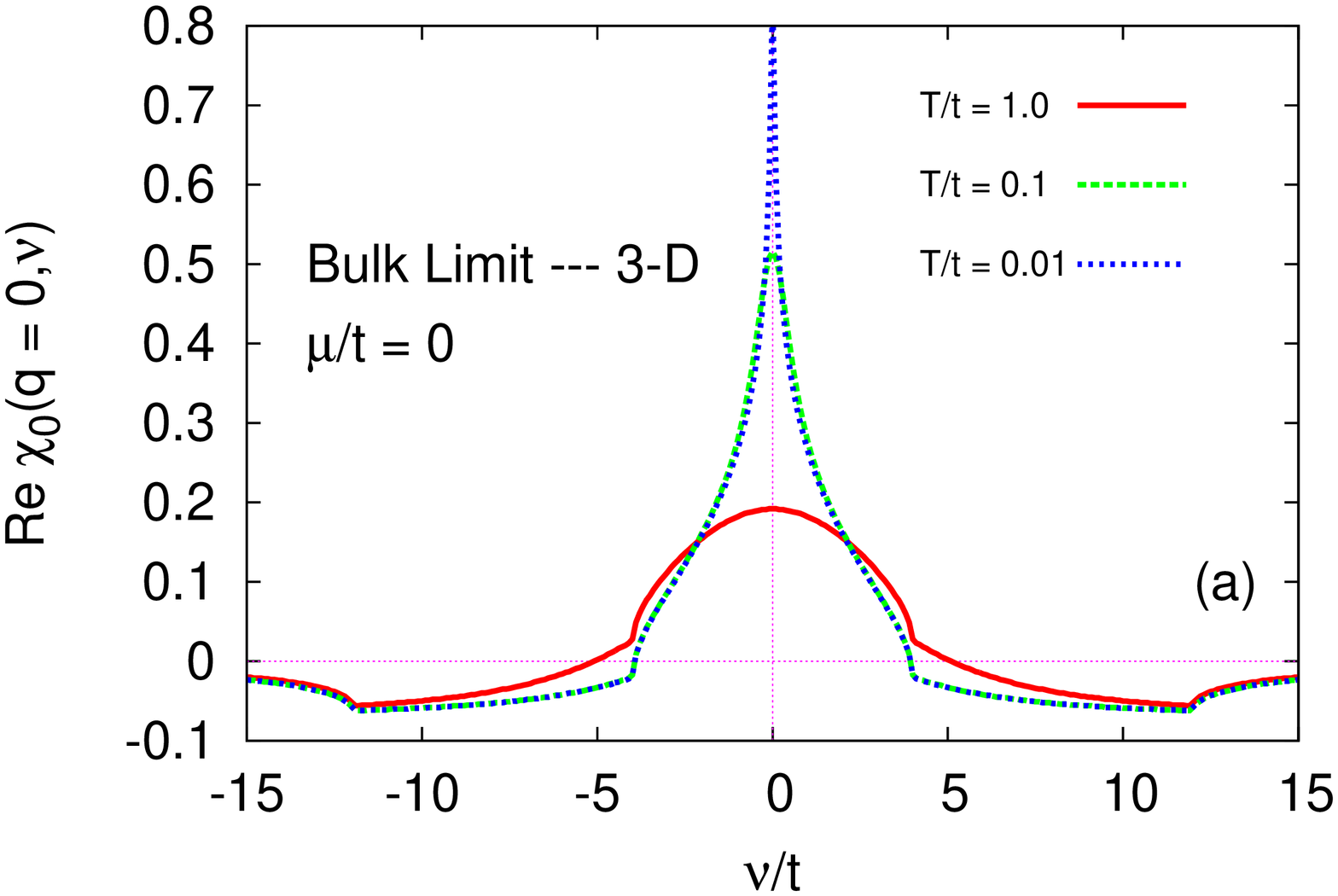}
\includegraphics[height=9cm,width=5cm,angle=-90,scale=0.9]{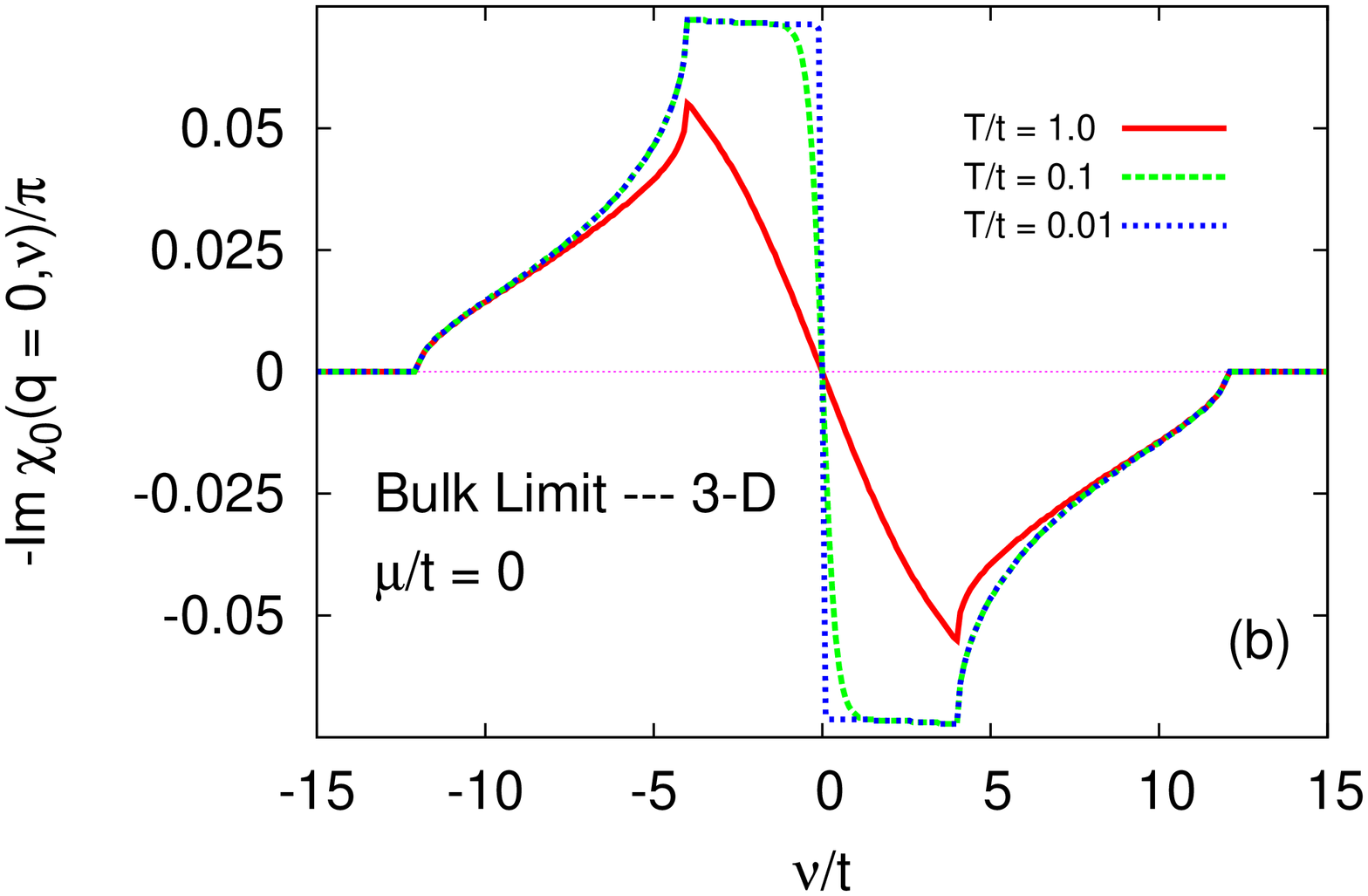}
\caption{Real (a) and Imaginary (b) parts of the noninteracting susceptibility
at zero wavevector vs.\ frequency, for three different temperatures, for the bulk
limit in three dimensions. In 3 dimensions no singularities occur at $\pm 12t$,
since the single electron density of states starts smoothly from zero. The logarithmic divergence 
in temperature remains at zero frequency.}
\label{fig3}
\end{center}
\end{figure}
\begin{figure}
\begin{center}
\includegraphics[height=9cm,width=5cm,angle=-90,scale=0.9]{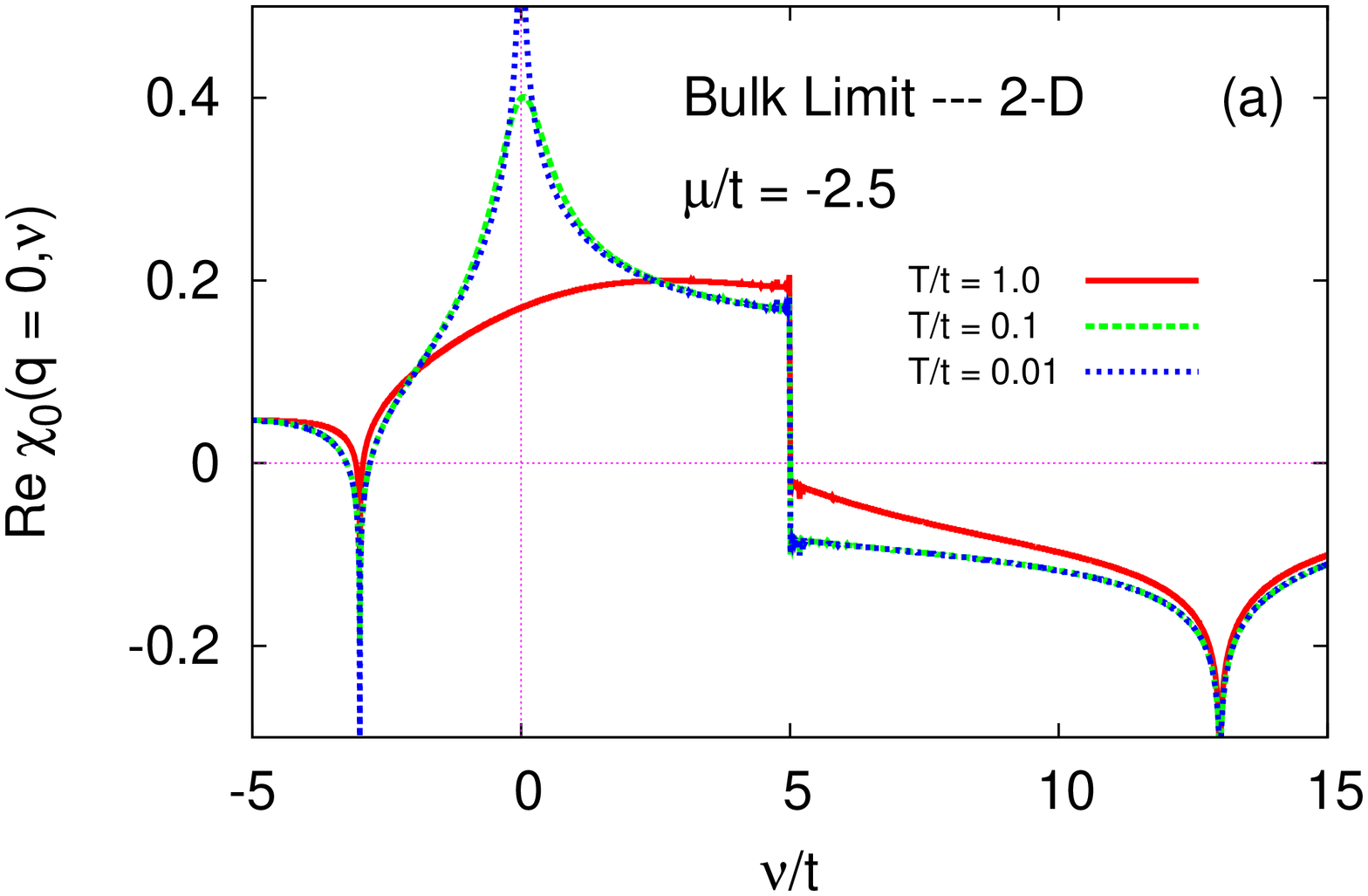}
\includegraphics[height=9cm,width=5cm,angle=-90,scale=0.9]{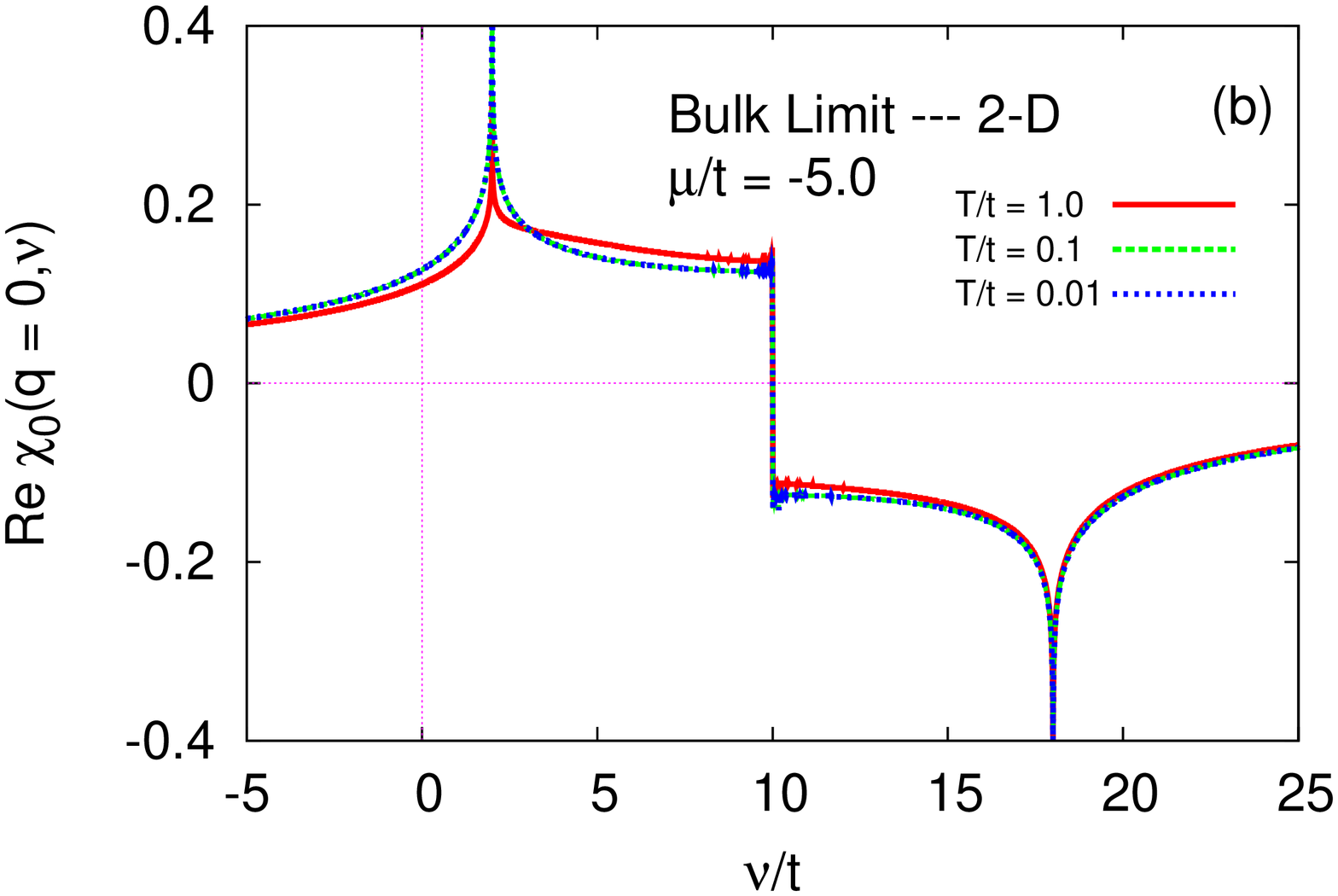}
\caption{Real part of the noninteracting susceptibility
at zero wavevector vs.\ frequency, for three different temperatures, for the bulk
limit in two dimensions, with nonzero chemical potential (a) above the bottom of the band, and (b) below the bottom of
the band. The chemical potential in (a)
($\mu = -2.5 t$) corresponds to some low filling of the band. Note that the divergence
in the real part occurs again at $\nu = 0$, while the imaginary part of the
susceptibility remains pinned to zero at this frequency. In contrast, the chemical potential in (b) ($\mu = -5.0 t$)  lies below the bottom
of the band; this means that the real part of the susceptibility diverges at all temperatures, and signals a bound
state being formed, even by two electrons. In 3D this does {\em not} occur for a single pair.}
\label{fig4}
\end{center}
\end{figure}

As mentioned, this calculation can be done in any dimension, and Fig.~(3a) and 
(3b) show the real and imaginary parts of the pair propagator in three
dimensions, with nearest neighbour hopping only, at half filling. Once again the essential feature is
that the real part diverges at zero frequency with decreasing temperature; thus, within mean field
theory, Eq.~\eqref{pairq} guarantees a transition when the real part crosses $1/|U|$. Of special note
is the lack of a negative divergence at the band edges in three dimensions; this affects the extreme dilute limit, and we will comment further
below.

Finally, while half-filling gives rise to symmetric results, more often than not some other instability intervenes to suppress
the pairing instability in this case. A result illustrating the pairing instability at a chemical potential away from half-filling is 
shown in Fig.~4, where the real part of the susceptibility is shown in two dimensions for (a) $\mu = -2.5t$ and (b) $\mu=-5.0$.
Note that the divergence in the 
real part persists in (a) at $\nu = 0$ as the temperature is lowered. The 
negative divergences associated with the band edge discontinuities (in 1D and 2D)
remain, but at a frequency $\pm W - 2 \mu$.
The discontinuity at $\nu = -2 \mu$ ($5t$ for this case) is only present 
here because of the logarithmic divergence in the 2D density of states 
at the origin and is absent for other band structures that lack 
singularities.  In (b) the chemical potential is below the bottom of the band; now there is a {\em positive} divergence at $\nu = -W - 2\mu$
(here at $2t$). This divergence is peculiar to 2D and 1D only and shows that any $|U|$, no matter how small, will lead to an instability in
the extreme dilute limit. This is {\em not} the case in higher dimension. For 3D, there is no band edge divergence, and it is well-known 
that an attractive potential must exceed some threshold before it will support a bound state for two particles.

Results in other dimensions away from half-filling are similar, albeit with differences reflecting the different densities of states, as already noted
at half-filling, and the critical difference just noted regarding band edge divergences in 3D vs.\ 2D and 1D. Furthermore, completely symmetric results 
occur for $\mu = +2.5$ (compared to $\mu = -2.5$), due to 
the particle-hole symmetry in the
problem. 

In summary we have computed the two particle pairing susceptibility in the thermodynamic limit, in a variety of dimensions
and for any filling. We have shown how the BCS instability comes about with decreasing temperature and how the nature of
the instability more closely resembles the one usually discussed in the context of mean-field ferromagnetism. Technically, at the instability temperature,
a single pole passes
from the lower half-plane to the upper-half-plane in complex frequency. This is in contrast to the finite lattice result, where two real roots become two pure imaginary roots, one of which leads to the instability. In either case the change at the instability signals a two-particle propagator that diverges with increasing time. 

The calculations in the thermodynamic limit enable one to see the dependence on dimensionality. Some quantitative differences occur due to the 
very different single-particle densities of states in 1D, 2D and 3D. However, in the extreme dilute limit, the different physics in 3D vs.\ 1D and 2D is highlighted
in the thermodynamic limit, and leads, in a natural way, to the necessity in 3D for Cooper's famous calculation.\cite{cooper56}

\begin{acknowledgments}

This work was supported in part by the Natural Sciences and Engineering
Research Council of Canada (NSERC), and by the Canadian
Institute for Advanced Research (CIfAR).

\end{acknowledgments}




\newpage

\begin{thebibliography}{99}

\bibitem{bardeen57}
J. Bardeen, L.N. Cooper and J.R. Schrieffer, Phys. Rev. {\bf 106}, 162 (1957);
Phys. Rev. {\bf 108}, 1175 (1957).

\bibitem{schrieffer64}
J.R. Schrieffer, {\it Theory of Superconductivity} (Benjamin/Cummings,
Don Mills, 1964).

\bibitem{rickayzen65}
G. Rickayzen, {\it Theory of Superconductivity} (John Wiley and Sons, Inc. New York, 1965).

\bibitem{thouless60} 
D.J. Thouless, Ann. Phys. {\bf 10}, 553 (1960). See also D.J. Thouless, {\it The Quantum Mechanics of Many-Body Systems},
(Academic Press, New York, 1961).

\bibitem{ambegaokar69}
V. Ambegaokar,  in {\it Superconductivity},
edited by R.D. Parks (Marcel Dekker, New York (1969)) Vol. 1, p.259.

\bibitem{timusk99} 
T. Timusk and B. Statt, Rep. Prog. Phys. {\bf 62}, 61 (1999).

\bibitem{pairing_review} This perhaps started with Thouless,\cite{thouless60} and was continued through the 1960's with Kadanoff and Baym and coworkers
(L.P. Kadanoff and P.C. Martin, Phys. Rev {\bf 124}, 670 (1961), G. Baym and L.P. Kadanoff, Phys. Rev. {\bf 124}, 287 (1961), and G. Baym, Phys. Rev. {\bf 127} 1391 (1962)). Eagles (D. M. Eagles, Phys. Rev. 186, 456 (1969)), Leggett (A.J. Leggett, J. de Physique, C7, {\bf 41}, 19 (1980); A.J. Leggett, in {\it Modern Trends in the Theory of Condensed Matter},
edited by S. Pekalski and J. Przystawa (Springer, Berlin, 1980)p. 13. ) and eventually  P. Nozi\`eres and S. Schmitt-Rink, J. Low Temp. Phys.
{\bf 59}, 195 (1985), addressed the same issues, but posed the problem as a weak-strong (BCS-BEC) cross-over issue, and how best to capture
effects omitted in the extreme (BCS or BEC) approaches. 

\bibitem{gooding04} References to work throughout the 1990's can be found in  R.J. Gooding, F. Marsiglio, S. Verga, and K.S.D. Beach, J. Low Temp. Phys. {\bf 136}, 191 (2004).

\bibitem{chen05} A review with applications to cold atom lattices can be found in Q.J. Chen, J. Stajic, S. Tan, and K. Levin,
Phys. Rep. {\bf 412}, 1 (2005).

\bibitem{doniach74} For example, with the Stoner criterion for magnetism; see, for example, S. Doniach and E.H. Sondheimer, {\it Green's Functions
for Solid State Physicists}, (Benjamin/Cummings, Don Mills, 1974).

\bibitem{beach01} For notation, etc. see, for example, K.S.D. Beach, R.J. Gooding, and F. Marsiglio, Phys. Lett. A
{\bf 282}, 319 (2001), and, more recently, S. Verga, R.J. Gooding, and F. Marsiglio, Phys. Rev. B
{\bf 71}, 155111 (2005).


\bibitem{vilk97} This is most emphasized in the two-particle self-consistent approach of J. Vilk and A.-M.S. Tremblay, J. de Physique I (France)
{\bf 7}, 1309 (1997). See also A.-M.S. Tremblay, B. Kyung and D. Senechel, Low. Temp. Phys. {\bf 32}, 424 (2006); [Fiz. Nizk. Temp. 32, 561 (2006)].

\bibitem{kadanoff61} See also Kadanoff and Martin in Ref. [\onlinecite{pairing_review}].

\bibitem{remark} The single-particle densities of states are readily determined ahead of time for a hypercubic lattice
in one, two, and three dimensions. See, for example, Appendix A in F. Marsiglio and J.E. Hirsch, Phys. Rev. B{\bf 41}, 6435 (1990).

\bibitem{cooper56} L.N. Cooper, Phys. Rev. {\bf 104}, 1089 (1956).

\end{thebibliography}
\end{document}